\documentclass[preprint,showkeys,secnumarabic,amsfonts,showpacs,amsmath,amssymb]{revtex4}
\usepackage[dvips]{color}
\usepackage{epsfig}
\usepackage{amsmath}
\usepackage{graphicx}
\def\Box{\hbox{$\rlap{$\sqcup$}\sqcap$}}

\begin{document}
\title{Bouncing Universe and phantom crossing in Modified Gravity and its reconstruction}

\author{Hossein Farajollahi}
\email{hosseinf@guilan.ac.ir} \affiliation{Department of Physics,
University of Guilan, Rasht, Iran}
\author{Farzad Milani}
\email{fmilani@guilan.ac.ir} \affiliation{Department of Physics,
University of Guilan, Rasht, Iran}

\date{\today}

\begin{abstract}
\noindent \hspace{0.35cm} In this paper we consider FRW cosmology in modified gravity which contain
arbitrary functions $f(\phi)$. It is shown that the bouncing solution appears in
the model whereas the equation of state (EoS) parameter crosses the phantom divider.
The reconstruction of the model is also investigated with the aim to
reconstruct the arbitrary functions and variables of the model.\\

\end{abstract}



\maketitle

\section{Introduction}

There are many cosmological observations, such
as Super-Nova Ia (SNIa) {\cite{c14-02}}, Wilkinson Microwave
Anisotropy Probe (WMAP) {\cite{c15}}, Sloan Digital Sky Survey
(SDSS) {\cite{c16-02}}, Chandra X-ray Observatory
{\cite{c17}} etc., that reveal some cross-checked information
of our universe. They suggest that the universe is spatially flat,
and consists of approximately $70\%$ dark energy (DE) with negative
pressure, $30\%$ dust matter (cold dark matters plus baryons), and
negligible radiation, and also the universe is undergoing an
accelerated expansion.

Recent observations have determined basic cosmological parameters in
high-precisions, but at the same time they posed a serious problem
about the origin of DE. The combined analysis of
 SNIa {\cite{a1-02}}, that is based upon the
background expansion history of the universe around the redshift $z
< \mathcal{O}(1)$, galaxy clusters measurements and WMAP data,
provides an evidence for the accelerated cosmic expansion
{\cite{c3-02}}. The cosmological acceleration strongly
indicates that the present day universe is dominated by smoothly
distributed slowly varying DE component. The constraint obtained
from SNIa so far has a degeneracy in the EoS of
DE {\cite{a2-03}}. To many people's frustration, the
$\Lambda\text{CDM}$ model with an EoS $\omega = -1$ has been
continuously favored from observations. This degeneracy has been
present even by adding other constraints coming from Cosmic
Microwave Background (CMB) {\cite{a3-02}} and Baryon Acoustic
Oscillations (BAO) {\cite{a4}}. The modern constraints on the EoS
parameter are around the cosmological constant value, $\omega
= -1 \pm 0.1$ {\cite{c3-02}}-{\cite{c4-02}} and a possibility that
$\omega$ is varied in time is not excluded. From the theoretical
point of view there are three essentially different cases:
$\omega>-1$ (quintessence), $\omega = -1$ (cosmological constant)
and $\omega <
-1$ (phantom) ({\cite{c5-06}}-{\cite{c8-03}} and refs. therein).

The models of DE can be broadly classified into two classes
{\cite{a5,a6}}. The first corresponds to introducing a specific
matter that leads to an accelerated expansion. Most of scalar field
models such as quintessence {\cite{a7-05}} and
k-essence {\cite{a8-02}} belong to this class. The second
class, that in this paper we consider, corresponds to the so-called
modified gravity models such as $f(R)$ gravity
{\cite{a9-03}}, scalar-tensor theories
{\cite{a10-11}} and brane-world models
{\cite{a11-02}}. In order to break the degeneracy of
observational constraints on $\omega$ and to discriminate between a
DE models, it is important to find additional information
other than the background expansion history of the Universe
{\cite{a12}}.

In second classification, modified gravity {\cite{b1}} suggests fine
alternative for DE origin. Indeed, it may be naturally expected that
gravitational action contains some extra terms which became relevant
recently with the significant decrease of the universe curvature.
The modified gravity can be obtained in two ways, first by replacing
scalar curvature $R$, or with $f(R)$, in the action which is well
known as modified gravity ,or $f(R)$  modified gravity, and second
by considering additional curvature invariant terms like
Gauss-Bonnet (GB) term. Another modification of $GR$ is modified
Gauss-Bonnet gravity {\cite{prd}} which is obtained by inserting a
function of GB invariant $f(G)$, in the Einstein-Hilbert action. A
number of metric formulation of modified $f(R)$ gravities has been
proposed {\cite{b1}}-{\cite{b5-12}} which explain the origin of
cosmic acceleration. Particular attention is paid to $f(R)$ models
{\cite{b6-03}}-{\cite{b9}} with the effective cosmological constant
phase because such theories may easily reproduce the well-known
$\Lambda\text{CDM}$ cosmology. Such models subclass which does not
violate Solar System tests represents the real alternative for
standard General Relativity.{\cite{b10}}

On the other hand, the Friedman equation forms the starting point
for almost all investigations in cosmology. Over the past few years
possible corrections to the Friedman equation have been derived or
proposed in a number of different contexts, generally inspired by
brane-world investigation {\cite{c9, c10}}. These modification are
often of a form that involves the total energy density $\rho$. In
{\cite{c11}}, multi-scalar coupled to gravity is studied in the
context of conventional Friedman cosmology. It is found that the
cosmological trajectories can be viewed as geodesic motion in an
augmented target space.

There are several phenomenological models describing the crossing of
the cosmological constant barrier {\cite{c12-05, c13-04}}.
Most of them use more then one scalar field or use a non-minimal
coupling with the gravity, or modified gravity, however we use both
of them in our paper. In two-field models one of these two fields is
a phantom, other one is a usual field and the interaction is
non-polynomial in general. It is important to find a model which
follows from the fundamental principles and describes a crossing of the $\omega = -1$ barrier.

A bouncing universe which provides a possible solution to the Big Bang singularity problem in standard cosmology has recently attracted a lot of interest in the field of string
theory and modified gravity {\cite{Biswas, Kanti}}. In bouncing cosmology,  within the framework of the standard FRW cosmology the null energy condition (NEC) for a period of time around the
bouncing point is violated . Moreover, after the bouncing when the universe enters into the hot Big Bang era, the EoS parameter $\omega$
in the universe transits from $\omega <-1$ to $\omega >-1$ \cite{Cai}.

In this paper, in section 2 we study the dynamics of the FRW cosmology in modified (non-local) gravity.
We discuss analytically and numerically a detailed examination of the conditions for having
$\omega$ across over $-1$. The necessary conditions
required for a successful bounce is discussed in this section as well.
Section 3 describes our model reconstruction and present the results.
Finally, we summaries our paper in section 4.

\section{The Model}

We start with the action of the non-local gravity as a simple
modified garvity given by {\cite{b10}},
\begin{eqnarray}\label{ac1}
S=\int{d^4x\sqrt{-g}\left\{\frac{M^{2}_{p}}{2}R(1+f(\Box^{-1}R))\right\}}\cdot
\end{eqnarray}
where $M_{p}$ is Plank mass, $f$ is some function and $\Box$ is
d'Almbertian for scalar field. Generally speaking, such non-local
effective action, derived from string theory, may be induced by quantum effects. A Bi-
scalar reformation of non-local action can be presented by introducing two scalar
fields $\phi$ and $\psi$, where changes the above action to a local
from:
\begin{eqnarray}
S&=&\int{d^{4}x\sqrt{-g}\left\{\frac{M^{2}_{p}}{2}\left[R(1+f(\phi))+\psi(\Box\phi-R)\right]\right\}},\label{ac2}
\label{ac}
\end{eqnarray}
where $\psi$, at this stage, plays role of a lagrange multiplier. One might
further  rewrite the above action as
\begin{eqnarray}
S&=&\int{d^{4}x\sqrt{-g}\left\{\frac{M^{2}_{p}}{2}\left[R(1+f(\phi)-\psi)-\partial_
{\mu}\psi\partial^{\mu}\phi\right]\right\}},\cdot\label{ac}
\end{eqnarray}
which now is equivalent to a local model with two extra degrees of
freedom. By the variation over $\psi$, we obtain $\Box\phi=R$ or
$\phi= \Box^{-1}R$, where $f$ is a function of scalar field
 $\phi$ in the model.

  Now in a FRW cosmological model with
 invariance of the action under changing fields and
vanishing variations at the boundary, the equations of motion for
only time dependent scalar fields, $\phi$ and $\psi$,  become
\begin{eqnarray}
\ddot{\phi}+3H\dot{\phi}+R&=&0,\label{EoS1}\\
\ddot{\psi}+3H\dot{\psi}-Rf'&=&0,\label{EoS2}
\end{eqnarray}
where $R=12H^2+6\dot{H}$,  $H$ is Hubble parameter and $f'=\frac{df(\phi)}{d\phi}$.
Variation of action (\ref{ac}) with respect to the metric tensor $g_{\mu\nu}$ gives,
\begin{eqnarray}\label{T}
0&=&\frac{1}{2}g_{\mu\nu}\left\{R(1+f-\psi)-\partial_{\rho}\psi\partial^{\rho}\phi\right\}-R_{\mu\nu}(1+f-\psi)
+\frac{1}{2}(\partial_{\mu}\psi\partial_{\nu}\phi+\partial_{\mu}\phi\partial_{\nu}\psi)\nonumber\\
&-&(g_{\mu\nu}\Box-\nabla_{\mu}\nabla_{\nu})(f-\psi)\cdot
\end{eqnarray}
The $00$ and $ii$ components of the equation (\ref{T}) are
 \begin{eqnarray}
0&=&-3H^2(1+f-\psi)+\frac{1}{2}\dot{\psi}\dot{\phi}-3H(f'\dot{\phi}-\dot{\psi}),\label{f1}\\
0&=&(2\dot{H}+3H^2)(1+f-\psi)+\frac{1}{2}\dot{\psi}\dot{\phi}+(\frac{d^2}{dt^2}+2H\frac{d}{dt})(f-\psi)\cdot\label{f2}
\end{eqnarray}
Equations (\ref{f1}) and (\ref{f2}) can be rewritten as
\begin{eqnarray}
3H^2&=&\frac{\frac{1}{2}\dot{\psi}\dot{\phi}-3H(f'\dot{\phi}-\dot{\psi})}{(1+f-\psi)},\label{f3}\\
2\dot{H}+3H^2&=&-\frac{\frac{1}{2}\dot{\psi}\dot{\phi}+(\frac{d^2}{dt^2}+2H\frac{d}{dt})(f-\psi)}{(1+f-\psi)}\cdot\label{f4}
\end{eqnarray}
Comparison with the standard Friedman equations
$H^{2}=\frac{\rho_{eff}}{3M_{p}^{2}}$, and $2\dot{H}+3H^2=-\frac{p_{eff}}{2M_{p}^{2}}$,
the right hand side of the equations (\ref{f3}) and (\ref{f4}) can be treated as the effective energy density and pressure:
\begin{eqnarray}
\frac{\rho_{eff}}{M_{p}^{2}}&=&\frac{\frac{1}{2}\dot{\psi}\dot{\phi}-
3H(f'\dot{\phi}-\dot{\psi})}{(1+f-\psi)},\label{f5}\\
\frac{p_{eff}}{M_{p}^{2}}&=&\frac{\frac{1}{2}\dot{\psi}\dot{\phi}+(\frac{d^2}{dt^2}+2H\frac{d}{dt})
(f-\psi)}{(1+f-\psi)}\cdot\label{f6}
\end{eqnarray}
Using Eqs. (\ref{EoS1}) and (\ref{EoS2}) and doing some algebraic calculation we can read the effective
 energy density and pressure from the above as,
\begin{eqnarray}\label{rho}
\rho_{eff}=\frac{M^{2}_{p}}{1+f-\psi}\left\{\frac{1}{2}\dot{\psi}\dot{\phi}-3H(f'\dot{\phi}-\dot{\psi})\right\}\cdot
\end{eqnarray}
\begin{eqnarray}\label{p}
p_{eff}=\frac{M^{2}_{p}}{1+f-\psi-6f'}\left\{\frac{1}{2}
\dot{\psi}\dot{\phi}+f''\dot{\phi}^{2}-
H(f'\dot{\phi}-\dot{\psi})+\frac{f'\left[6H(f'\dot{\phi}-\dot{\psi})
-\dot{\psi}\dot{\phi}\right]}{1+f-\psi}\right\}\cdot
\end{eqnarray}
Now by using Eqs. (\ref{rho}) and (\ref{p}) the conservation
equation can be obtained as,
\begin{eqnarray}\label{rhodot}
\dot{\rho}_{eff}+3H\rho_{eff}(1+\omega)=0,
\end{eqnarray}
where $\omega=\frac{p_{eff}}{\rho_{eff}}$ is the EoS parameter of the model. Also from Eq. (\ref{f1}) we obtain,
\begin{eqnarray}\label{h}
H=\frac{-(f'\dot{\phi}-\dot{\psi})}{2(1+f-\psi)}
\left\{1\pm\sqrt{1+\frac{6}{9}\frac{\dot{\psi}\dot{\phi}
(1+f-\psi)}{(f'\dot{\phi}-\dot{\psi})^{2}}}\right\}\cdot
\end{eqnarray}
At this stage we study the cosmological evolution of EoS parameter, $\omega$ , and show that analytically and numerically there are conditions that
cause the EoS parameter crosses the phantom divide line ($\omega\rightarrow -1$).
Let's see under what conditions the system will be able to cross the barrier of $\omega = -1$. In order to do that, one requires $\rho_{eff} + p_{eff}$ to vanish at a point of $(\phi_0,\psi_0)$ and change the sign after the crossing. This can only be achieved by
requiring $\dot{H}(\phi_0,\psi_0) = 0$ and $\dot{H}$ has different signs before and after the crossing.

To explore this
possibility, we have to check the condition
$\frac{d}{dt}(\rho_{eff}+p_{eff})\neq 0$ when $\omega\rightarrow -1$. Using Eqs.
(\ref{rho}) and (\ref{p}) in second Friedman equation, Eq. (\ref{f4}) gives,
\begin{eqnarray}\label{Hdot}
\dot{H}=\frac{24f'H^{2}+
4H(f'\dot{\phi}-\dot{\psi})-(f''\dot{\phi}^{2}+\dot{\psi}\dot{\phi})}{2(1+f-\psi-6f')}\cdot
\end{eqnarray}
Also we have $\frac{d}{dt}(\rho_{eff}+p_{eff})=
-2M^{2}_{p}\ddot{H}$ or,
\begin{eqnarray}\label{hddot}
\ddot{H}&=&\frac{\ddot{\phi}(4Hf'-2f''\dot{\phi}-\dot{\psi})-\ddot{\psi}(4H+\dot{\phi})
+\dot{\phi}\Big(4Hf''(6H+\dot{\phi})-f'''\dot{\phi}^{2}\Big)}{2(1+f-\psi-6f')}\nonumber\\
&+&\dot{H}\left(\frac{24f'H+\dot{\phi}(f'+6f'')-\dot{\psi}}{1+f-\psi-6f'}\right)
\cdot
\end{eqnarray}
One can find that in order to have $\omega$-crossing one of the following conditions might be satisfied when $\omega\rightarrow -1$ and $H\neq 0$.
\begin{itemize}
  \item (a) $\dot{\psi}=0$ and $\dot{\phi}\neq 0$
  \item (b) $\dot{\phi}=0$ and $\dot{\psi}\neq 0$
  \item (c) $\ddot{\phi}=0$ and $\ddot{\psi}=0$
\end{itemize}
In the first case, (a),we have
\begin{eqnarray}\label{hb}
H=\frac{-f'\dot{\phi}}{1+f-\psi},
\end{eqnarray}
\begin{eqnarray}\label{Hdotb}
\dot{H}=\frac{4H(6f'H+\dot{\phi})-f''\dot{\phi}^{2}}{2(1+f-\psi-6f')},
\end{eqnarray}
and,
\begin{eqnarray}\label{hddotb}
\ddot{H}&=&\frac{2\ddot{\phi}
(2Hf'-f''\dot{\phi})-\ddot{\psi}(4H+\dot{\phi})
+\dot{\phi}\left(4Hf''(6H+\dot{\phi})-f'''\dot{\phi}^{2}\right)}{2(1+f-\psi-6f')}\cdot
\end{eqnarray}

Therefore the conditions for having $\omega$ across over $-1$ are: (a-1) $\ddot{\phi}\neq 0$ and $2Hf'\neq f''\dot{\phi}$ when other terms can be neglected, (a-2) $\ddot{\psi}\neq 0$ however the first and third terms can be vanished, (a-3) $2\ddot{\phi}(2Hf'-f''\dot{\phi})\neq \ddot{\psi}(4H+\dot{\phi})$ where third term is zero or (a-4) $4Hf''(6H+\dot{\phi})\neq f'''\dot{\phi}^{2}$ or if $f'''=0$ then $f''\neq 0$ and if $f''=0$ then $f'''\neq 0$, when the first and second terms can be vanished in addition to the $f''\dot{\phi}^2=4H(6Hf'+\dot{\phi})$ and $1+f-\psi\neq6f'$.

In the second case, (b), we have,
\begin{eqnarray}\label{ha}
H=\frac{\dot{\psi}}{1+f-\psi},
\end{eqnarray}
\begin{eqnarray}\label{Hdota}
\dot{H}=\frac{2H(6f'H-\dot{\psi})}{1+f-\psi-6f'},
\end{eqnarray}
and,
\begin{eqnarray}\label{hddota}
\ddot{H}&=&\frac{\ddot{\phi}
(4Hf'-\dot{\psi})-4H\ddot{\psi}}{2(1+f-\psi-6f')}\cdot
\end{eqnarray}

This case never occurs because when $\dot{H}=0$ and $H\neq0$ we obtain $\dot{\psi}=6f'H$. By replacing it in Eq. (\ref{ha}) one leads to $6f'= 1+f-\psi$ which contradict our primary finding.

Finally, for the third case, (c), we have, $H$
and $\dot{H}$ as Eqs. (\ref{h}) and (\ref{Hdot})  in addition to,
\begin{eqnarray}\label{hddotc}
\ddot{H}=\frac{\dot{\phi}\left(4Hf''(6H+\dot{\phi})-f'''\dot{\phi}^{2}\right)}{2(1+f-\psi-6f')}\cdot
\end{eqnarray}
Therefore the conditions are (c-1) $4Hf''(6H+\dot{\phi})\neq f'''\dot{\phi}^{2}$, (c-2) $f'''=0$ and $f''\neq 0$, (c-3) $f''=0$ and $f'''\neq 0$, (c-4) $\dot{\phi}\neq 0$, in addition to the $4Hf'(6H^2+\dot{\phi})-f''\dot{\phi}^2=\dot{\psi}(4H+\dot{\phi})$ and $1+f-\psi\neq6f'$.

With numerical calculation, as shown in Fig. 1, by appropriately choosing model parameters we construct a cosmological model so that crossing the phantom divide occurs at $t>0$ and crosses $-1$ around this point which is supported by observations \cite{c8-03}. In some sense this model is similar to Quintom dark energy models consisting of two quintessence and phantom fields \cite{Nozari}.

\begin{tabular*}{2.5 cm}{cc}
\includegraphics[scale=.3]{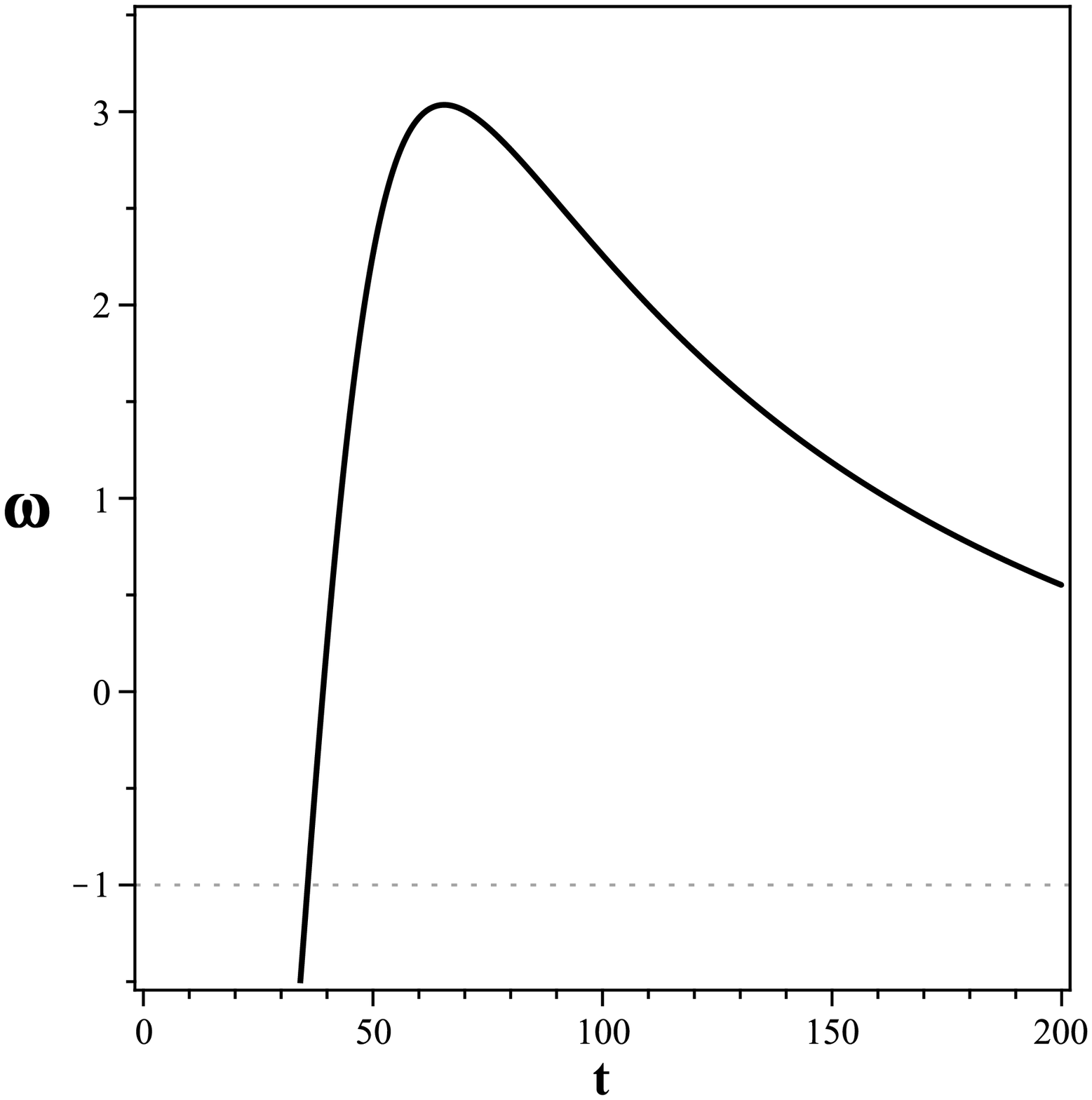}\hspace{0.2 cm}\\
\hspace{1 cm}Fig. 1: \, The graph of $\omega$ plotted as
function of time for ${\it f_{0}}\,{{\rm e}^{b\phi\left( t \right) }}$, $f_{0} = 0.5$ and\\
\hspace{.3 cm}  $b =1.5$. Initial values are $\phi(0)=0.5$ , $\dot{\phi}(0)=-0.02$, $\psi(0)=0.5$ , $\dot{\psi}(0)=0.01$.\\
\end{tabular*}\\

By choosing $t=0$ to be the bouncing point, the solution for $a(t)$ and $H(t)$, Eq. (\ref{h}), (see Fig. 2) provides a dynamical universe with
 contraction for $t<0$, bouncing at $t=0$ and then expansion for $t>0$.

\begin{tabular*}{2.5 cm}{cc}
\includegraphics[scale=.3]{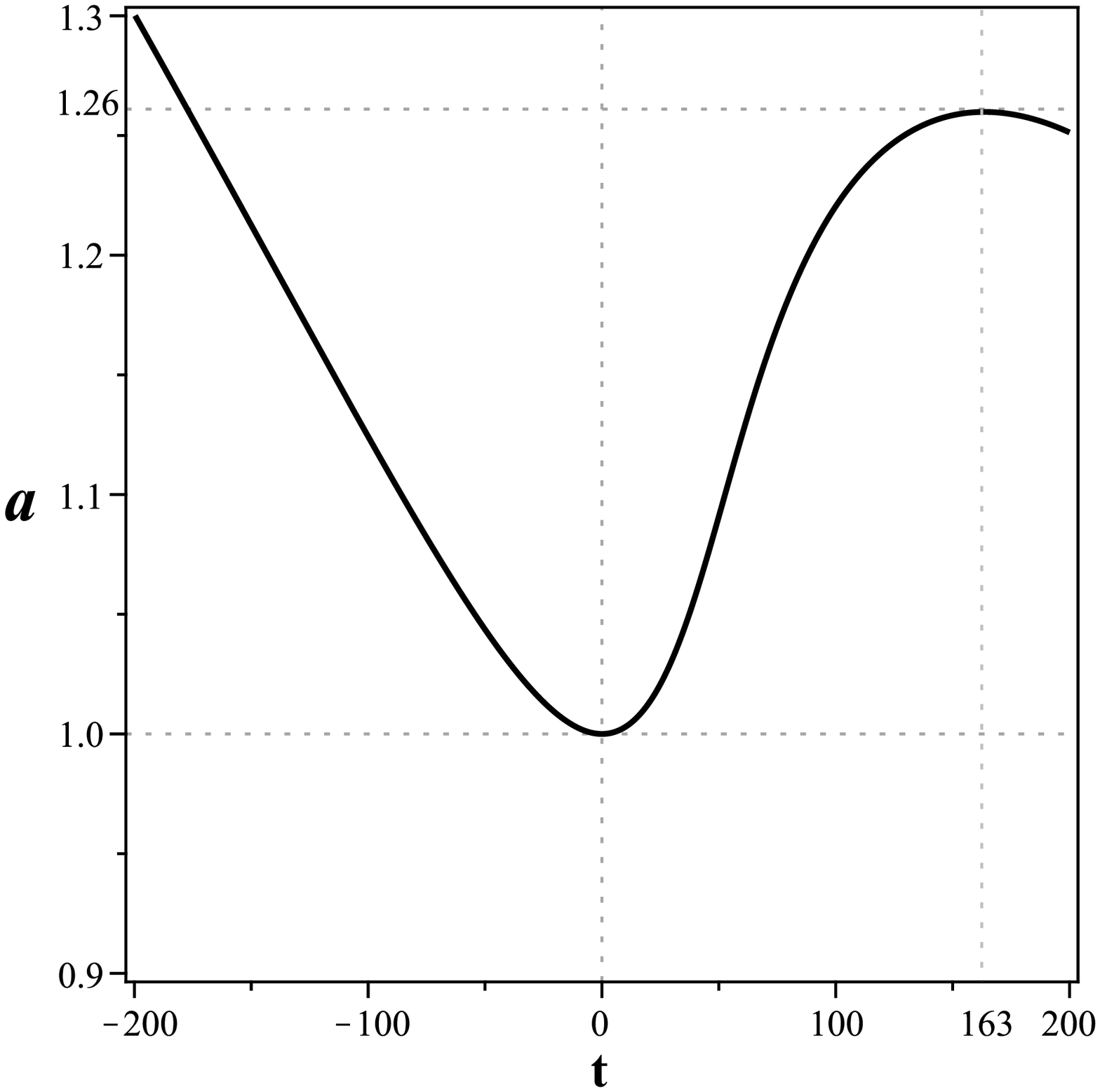}
\hspace{0.2 cm}\includegraphics[scale=.3]{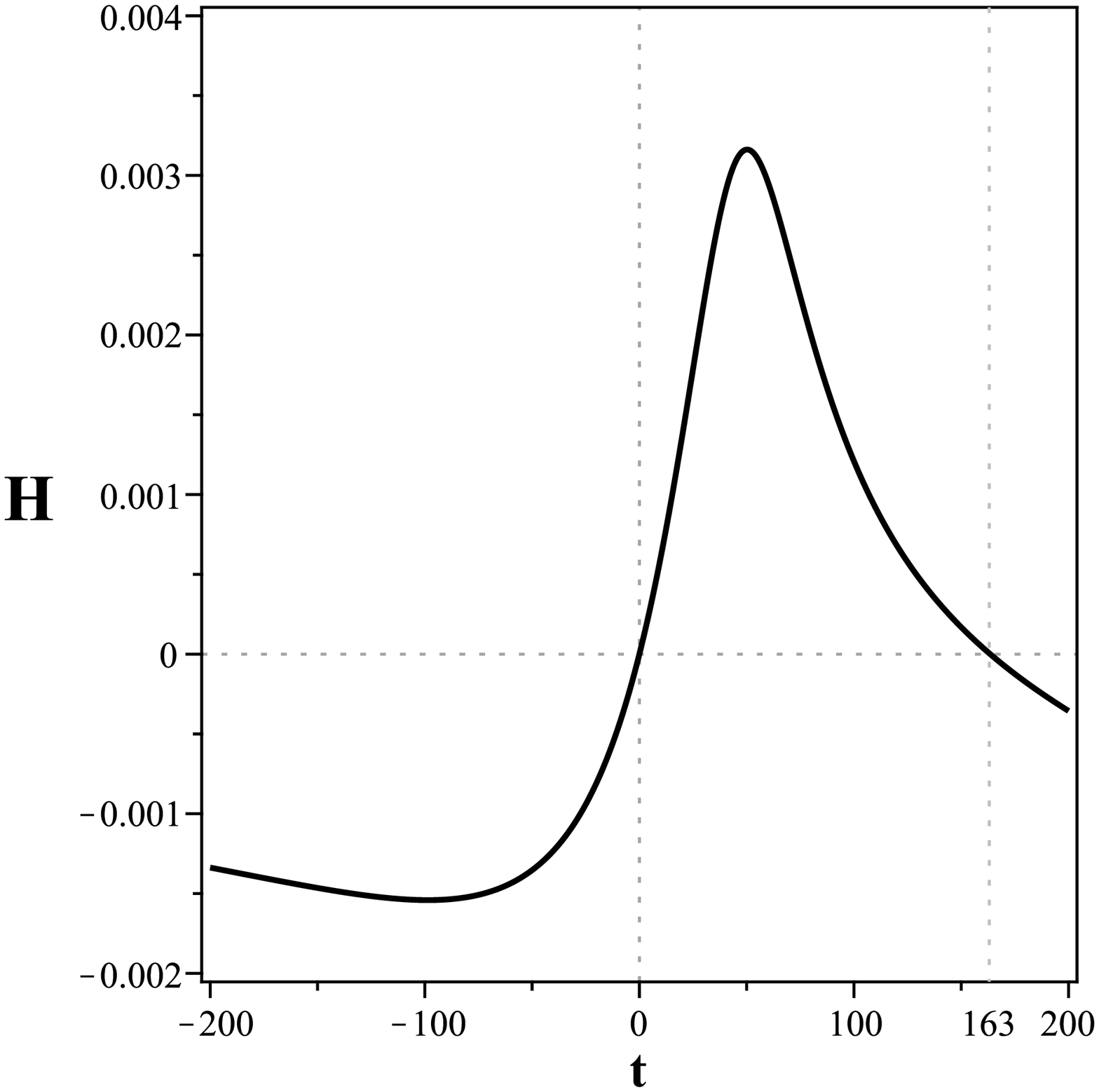}\\
\hspace{1 cm}Fig. 2: \, The graph of scale factor $a$ and $H$,  plotted as \\
\hspace{1 cm} function of time for ${\it f_{0}}\,{{\rm e}^{b\phi\left( t \right) }}$, $f_{0} = 0.5$, and $b = 1.5$.
 Initial values are\\
\hspace{1 cm}  $\phi(0)=0.5$ , $\dot{\phi}(0)=-0.02$, $\psi(0)=0.5$ , $\dot{\psi}(0)=0.01$.\\
\end{tabular*}\\

By definition, for a bounce or turn-around process to occur, one must require that at the
pivot point $\dot{a} = 0$ and $\ddot{a} > 0 $ around the bouncing point, while $\ddot{a} < 0 $ around the
turn-around point. According to Eqs. (\ref{f3}) and (\ref{f4}), for $\frac{\dot{\psi}\dot{\phi}}{2(1+f-\psi)}>0$ one can get
\begin{eqnarray}\label{bounce}
\rho_{eff}>0,\ \  p_{eff}<0 \ \ (\ \mbox{or}\ \  p_{eff}>0) \ \ \mbox{for the bounce (or turn-around)}
\end{eqnarray}
or equivalently, the EoS parameter approaches negative values (or positive values) at the bounce (or turn-around) point. This
shows $\omega$ possibly crosses over the cosmological constant boundary ($\omega= -1$), which interestingly implies the necessity to have the Quintom matter for the realization of the oscillating universe under modified gravity.

In the context of a spatially flat four-dimensional background FRW metric, if we are to obtain a smooth transition from a contracting universe into an expanding phase, there must be a period when NEC is violated . In this case, we need a kind of matter which admits an EoS parameter which is less
than -1, but only around the bounce. Neither regular not Phantom matter alone can achieve a transition
in the EoS parameter through the cosmological constant boundary. Therefore, a Quintom model is the
only possible solution to resolve this difficulty.

A detailed examination on the necessary conditions requires
for a successful bounce shows that during the contracting phase, the scale
factor $a(t)$ is decreasing, i.e., $\dot{a} < 0$, and in the
expanding phase we have $\dot{a} > 0$. At the bouncing point,
$\dot{a} = 0$, and so around this point $\ddot{a} > 0 $ for a
period of time. Equivalently in the bouncing cosmology the Hubble
parameter $H$ runs across zero from $H < 0$ to $H > 0$ and $H = 0$
at the bouncing point. A successful bounce requires that the following condition should be satisfied around bouncing
point,
\begin{eqnarray}\label{hdot1}
\dot{H}=-\frac{1}{2M^{2}_{p}}(1+\omega)\rho>0.
\end{eqnarray}
From Fig. 1  and 2, we see that at $t\rightarrow 0$, $\omega<-1$ and $\dot{H}$ is positive which satisfies the above
condition. Also we see that at the bouncing point where the scale factor $a(t)$ is not zero we avoid singularity
faced in the standard cosmology.

\section{Model Reconstruction}
Reconstructing of the model for the EoS parameter and deceleration parameter in three
forms of parametrization \cite{MPLA} has been studied in this section. From effective energy density and effective pressure, Eqs, (\ref{rho}) and
(\ref{p}) one can define construction function $\tilde{K}$ as:
\begin{eqnarray}\label{2rho}
3p_{eff}-\rho_{eff}=\frac{M^{2}_{p}}{1+f-\psi}(\dot{\psi}\dot{\phi}+6(\ddot{f}+3H\dot{f}))=2\tilde{K}\cdot
\end{eqnarray}
With comparison to Eq. (\ref{rho}) and using Eq. (\ref{h}) we have,
\begin{eqnarray}\label{rho-1}
\rho_{eff}=\tilde{K}-3\tilde{V},
\end{eqnarray}
where,
\begin{eqnarray}\label{vbar01}
\tilde{V}=\frac{M^{2}_{p}}{1+f-\psi}\left(\ddot{f}+4H\dot{f}-H\dot{\psi}\right)\cdot
\end{eqnarray}
One then simply finds that the energy pressure in terms of new functions $\tilde{K}$ an $\tilde{V}$ as,
\begin{eqnarray}\label{p-1}
p_{eff}=\tilde{K}-\tilde{V},
\end{eqnarray}
now we can rewrite the EoS parameter as,
\begin{eqnarray}\label{omega-1}
\omega=-1+\left(\frac{2\tilde{K}-4\tilde{V}}{\tilde{K}-3\tilde{V}}\right)\cdot
\end{eqnarray}
It can be seen that $\omega >-1$ when $\tilde{K}>3\tilde{V}$ or $\tilde{K}<2\tilde{V}$
and $\omega <-1$ when $2\tilde{V}<\tilde{K}<3\tilde{V}$ with the constrain that $\tilde{K}\neq 3\tilde{V}$.
The transition from $\omega>-1$ to $\omega<-1$ happens when $\tilde{K}=2\tilde{V}$.

In order to reconstruct the cosmological parameters we rewrite the modified Friedman equations in the present of dust matter as,
\begin{eqnarray}
3M^{2}_{p}H^{2}&=&\rho_{m}+\rho_{eff}=\rho_{m}+\tilde{K}-3\tilde{V},\label{3m}\\
-2M^{2}_{p}\dot{H}&=&\rho_{m}+\rho_{eff}+p_{eff}=\rho_{m}+2\tilde{K}-4\tilde{V},\label{2m}
\end{eqnarray}
where $\rho_{m}$ is the energy density of dust matter. We thus have new $\tilde{K}$ and $\tilde{V}$ in the present of matter:
\begin{eqnarray}
\tilde{K}&=&\frac{1}{2}\rho_{m}-3M^{2}_{p}(2H^{2}+\dot{H}),\label{kbar}\\
\tilde{V}&=&\frac{1}{2}\rho_{m}-M^{2}_{p}(3H^{2}+\dot{H})\cdot\label{vbar}
\end{eqnarray}
In here, we assume that the two effective and dust matter fluids do not interact. From \cite{c18-02} the expression for the energy density of dust
matter with respect to the redshift $z$ is given by,
\begin{eqnarray}\label{rhom}
\rho_{m}=3M^{2}_{p}H^{2}_{0}\Omega_{m0}(1+z)^{3},
\end{eqnarray}
where $\Omega_{m0}$ is the ratio density parameter of matter fluid
and the subscript $0$ indicates the present value of the
corresponding quantity. One then can rewrite $\tilde{K}$ and $\tilde{V}$
with respect to the redshift $z$ as,
\begin{eqnarray}
\tilde{K}&=&\frac{3}{2}M^{2}_{p}H^{2}_{0}\left(\Omega_{m0}(1+z)^{3}-4r+(1+z)r^{(1)}\right),\label{kbar1}\\
\tilde{V}&=&\frac{1}{2}M^{2}_{p}H^{2}_{0}\left(3\Omega_{m0}(1+z)^{3}-6r+(1+z)r^{(1)}\right),\label{vbar1}
\end{eqnarray}
where $r= \frac{H^{2}}{H^{2}_{0}}$ and
$r^{(n)}=\frac{d^{n}r}{dz^{n}}$. By using Eqs. (\ref{kbar1}) and (\ref{vbar1}) the EoS parameter can be rewritten as,
\begin{eqnarray}\label{omega-2}
\omega=\frac{(1+z)r^{(1)}-3r}{-3\Omega_{m0}(1+z)^{3}+3r},
\end{eqnarray}
where then $r(z)$ can be evaluated
\begin{eqnarray}\label{r}
r(z)=\Omega_{m0}(1+z)^{3}+(1-\Omega_{m0})e^{3\int^{z}_{0}\frac{1+\omega(\tilde{z})}{1+\tilde{z}}d\tilde{z}}\cdot
\end{eqnarray}
Moreover, by employing $\frac{d}{dt}=-(1+z)\frac{d}{dz}$, the deceleration parameter $q$ can be obtained as,
\begin{eqnarray}\label{q}
q(z)=-1-\frac{\dot{H}}{H^{2}}=\frac{(1+z)r^{(1)}-2r}{2r}\cdot
\end{eqnarray}
Now, with the following three different forms of parametrization, using numerical calculation, we reconstruct EoS and $q(z)$ parameters.\\

\textbf{Parametrization 1:}\\
This parametrization has been proposed by Chevallier and Polarski
\cite{c19} and Linder \cite{c20}, where the EoS parameter of dark
energy in term of redshift $z$ is given by,
\begin{eqnarray}\label{z1}
\omega(z)=\omega_{0}+\frac{\omega_{a}z}{1+z}\cdot
\end{eqnarray}
By fitting this model to the observational data we find that $\Omega_{m0}=0.29$, $\omega_{0}=-1.07$ and $\omega_{a}=0.85$ \cite{c23-02}

\textbf{Parametrization 2:}\\
The EoS parameter in term of redshift $z$ has been proposed by
Jassal, Bagla and Padmanabhan \cite{c21} as,
\begin{eqnarray}\label{z2}
\omega(z)=\omega_{0}+\frac{\omega_{b}z}{(1+z)^{2}}\cdot
\end{eqnarray}
where again fitting the data \cite{c23-02}, $\Omega_{m0}=0.28$, $\omega_{0}=-1.37$ and $\omega_{b}=3.39$

\textbf{Parametrization 3:} \\
The third parametrization has proposed by Alam, Sahni and Starobinsky
\cite{c22}, where $r(z)$ given by,
\begin{eqnarray}\label{z3}
r(z)=\Omega_{m0}(1+z)^{3}+A_{0}+A_{1}(1+z)+A_{2}(1+z)^{2}\cdot
\end{eqnarray}
This parametrization can be thought as the parametrization of $r(z)$ instead of $\omega(z)$.  By using the results in \cite{c23-02}, we get
coefficients  of the this parametrization as
$\Omega_{m0} = 0.30$,
$A_{0} = 1$, $A_{1} = -0.48$ and $A_{2} = 0.25$.\\

The evolution of $\omega(z)$ and $q(z)$ are plotted in Fig. 3 for the three form of parameterizations.
The figure shows that the EoS parameter crosses the phantom divide line for the first and second parametrization and never crosses the line in the third parametrization. The second parametrization crosses the phantom line in two different values of $z$.
As can be seen from the graph of deceleration parameter, for the second parametrization the universe undergoes
 an acceleration period from $z=0.39$ until now in comparison to the first and third parameterizations that acceleration starts earlier. \\

\begin{tabular*}{2.5 cm}{cc}
\includegraphics[scale=.3]{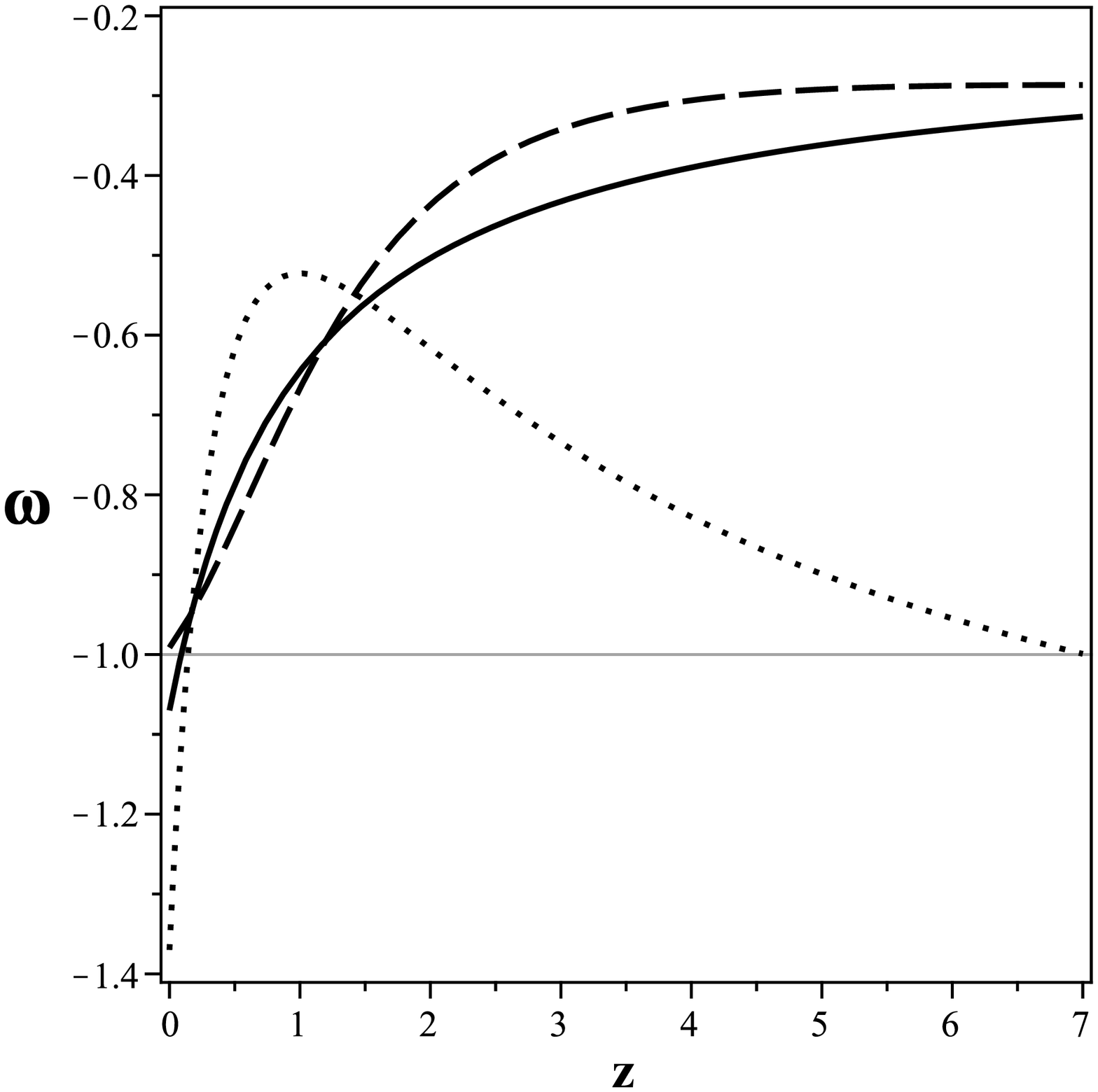} \hspace{0.5 cm} \includegraphics[scale=.3]{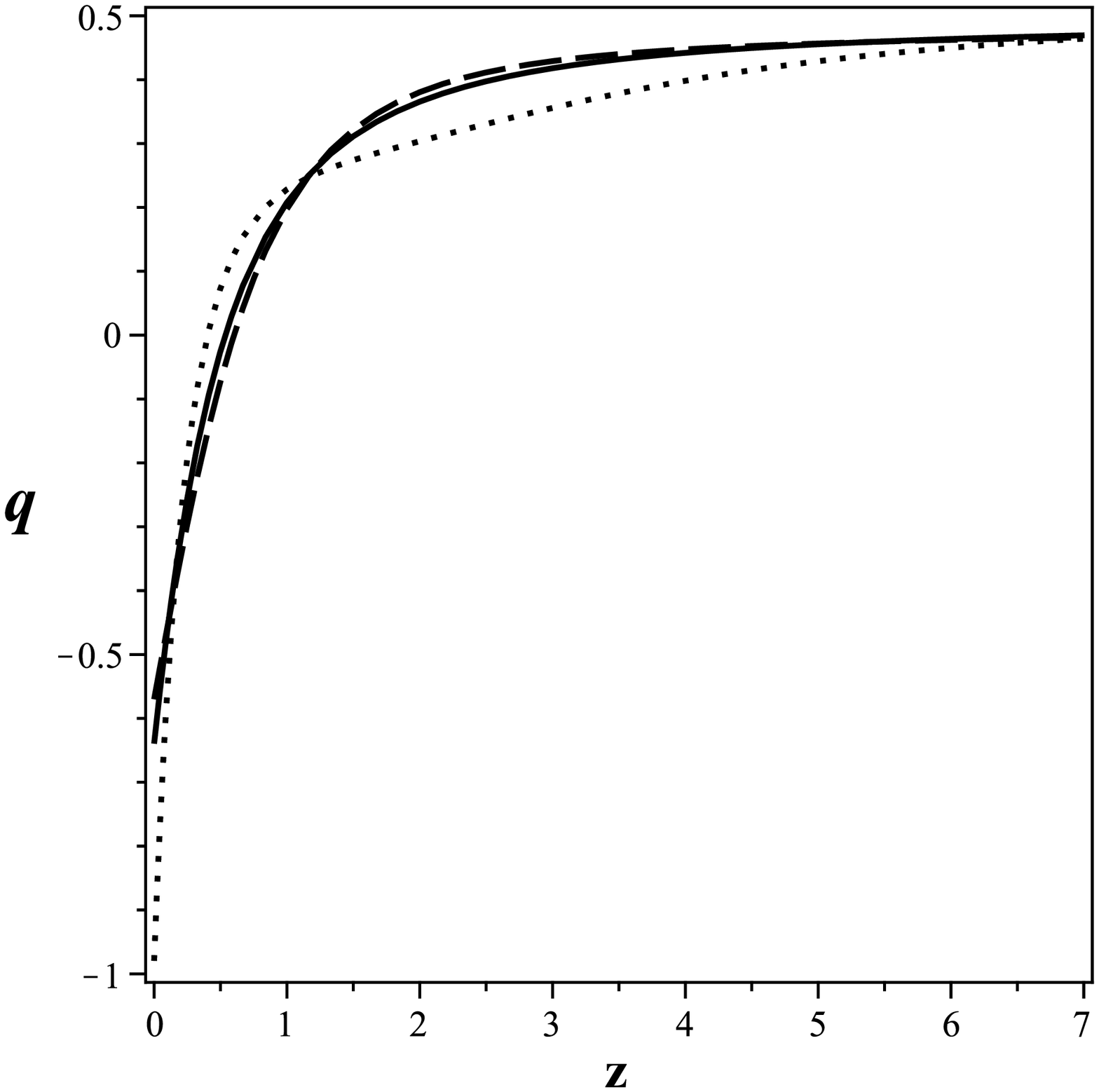}\\
\hspace{1cm}Fig. 3: \, The graphs of the EoS parameters, $\omega$, and deceleration parameters, $q$, with  \\
\hspace{1cm} respect to the redshift $z$. The solid, dot and dash lines represent parametrization 1,\\
\hspace{1cm} 2 and 3 respectively.\\
\end{tabular*}\\

Also, using Eqs. (\ref{kbar1}), (\ref{vbar1})
and the three parameterizations, the evolutions of $K (\tilde{z})$
and $V
(\tilde{z})$ are shown in Fig. 4.\\
\begin{tabular*}{2.5cm}{cc}
\includegraphics[scale=.3]{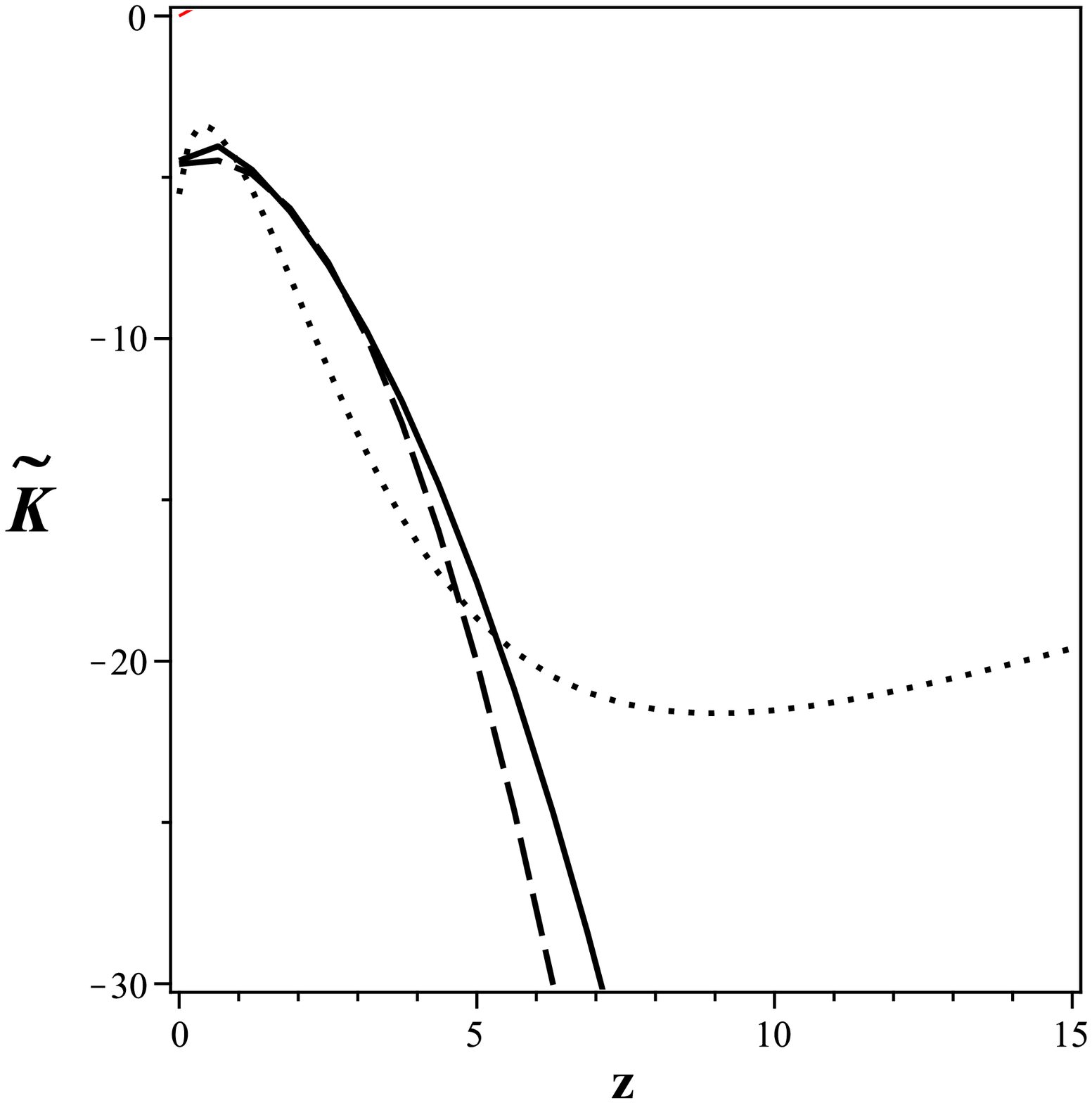} \hspace{1cm}\includegraphics[scale=.3]{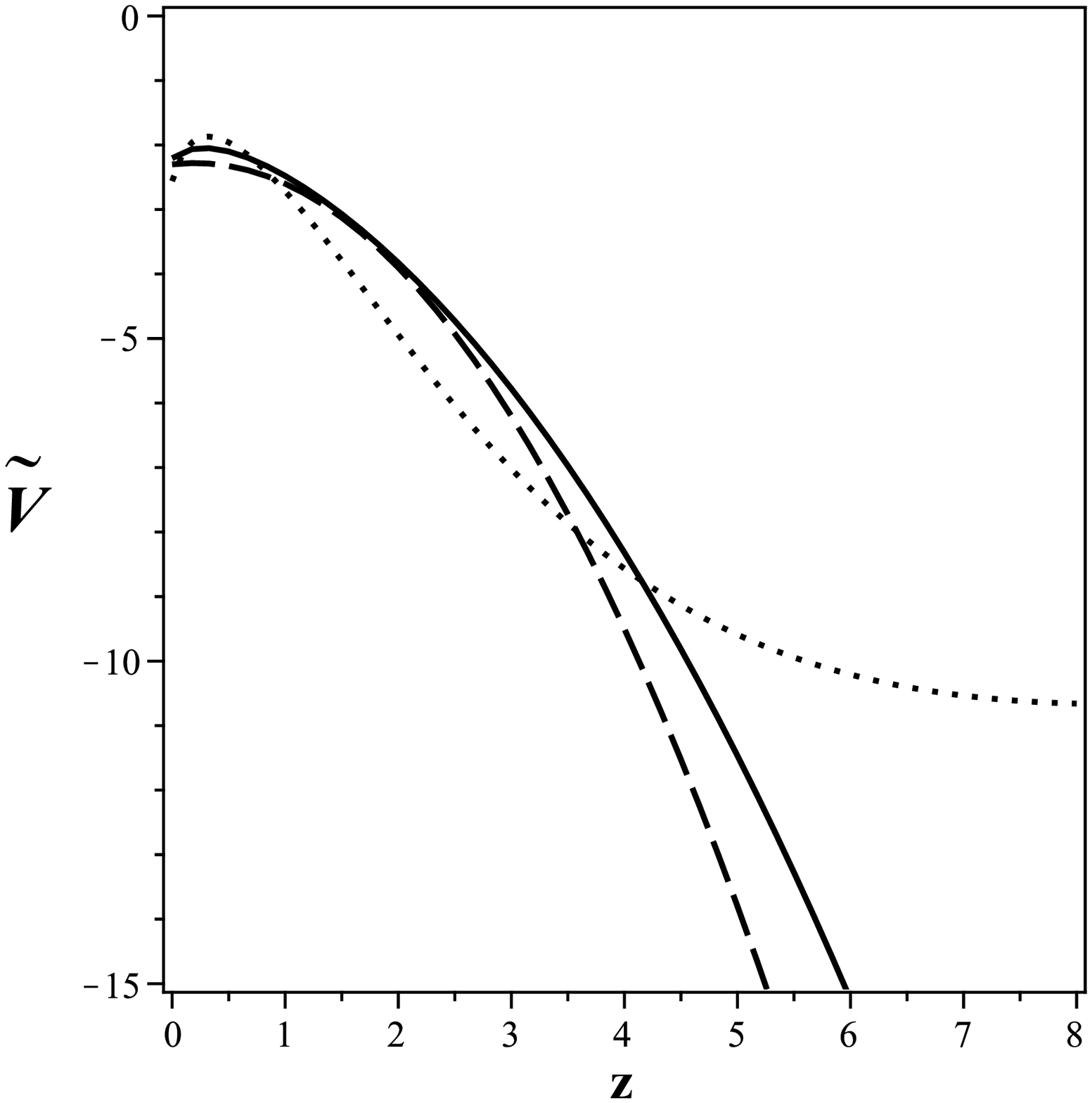}\\
\hspace{1cm}Fig. 4:\, The graphs of the reconstructed $\tilde{K}$ and $\tilde{V}$ with respect to the redshift $z$. The\\
\hspace{1cm} solid, dot and dash lines represent parametrization 1, 2 and 3 respectively.\\
\end{tabular*}\\

Since $\tilde{K}$ and $\tilde{V}$ are now known functions of $z$, we can obtain the evolutions of $\phi(z)$ and $f(z)$ with respect
of $z$, which are plotted in Fig. 5 for the three parameterizations.  One may also directly obtain the relationship between
the function $f$ and the scalar field $\phi$, which is plotted in Fig. 6 for the three parameterizations.\\

\begin{tabular*}{2.5 cm}{cc}
\includegraphics[scale=.3]{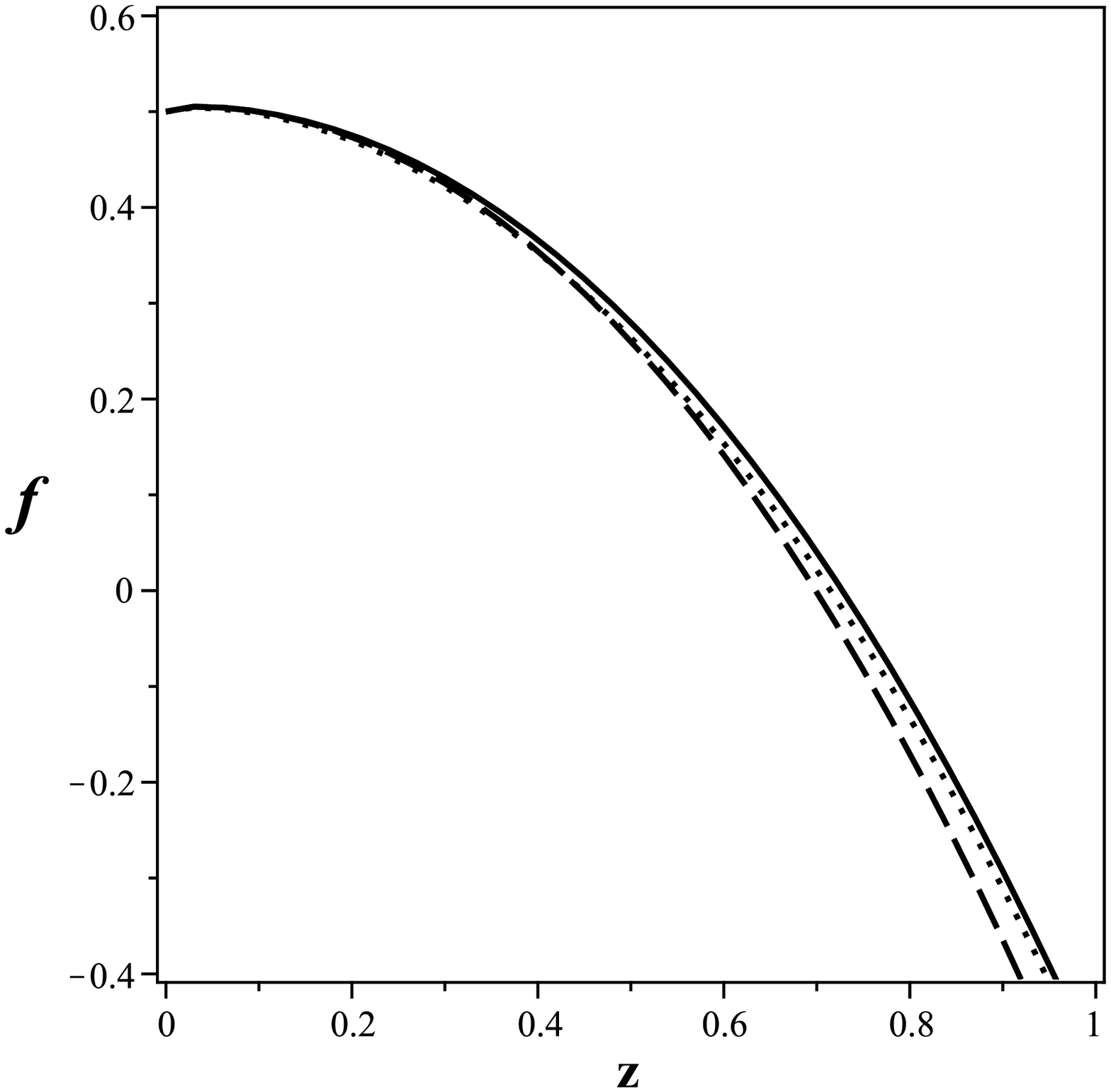} \hspace{1cm}\includegraphics[scale=.3]{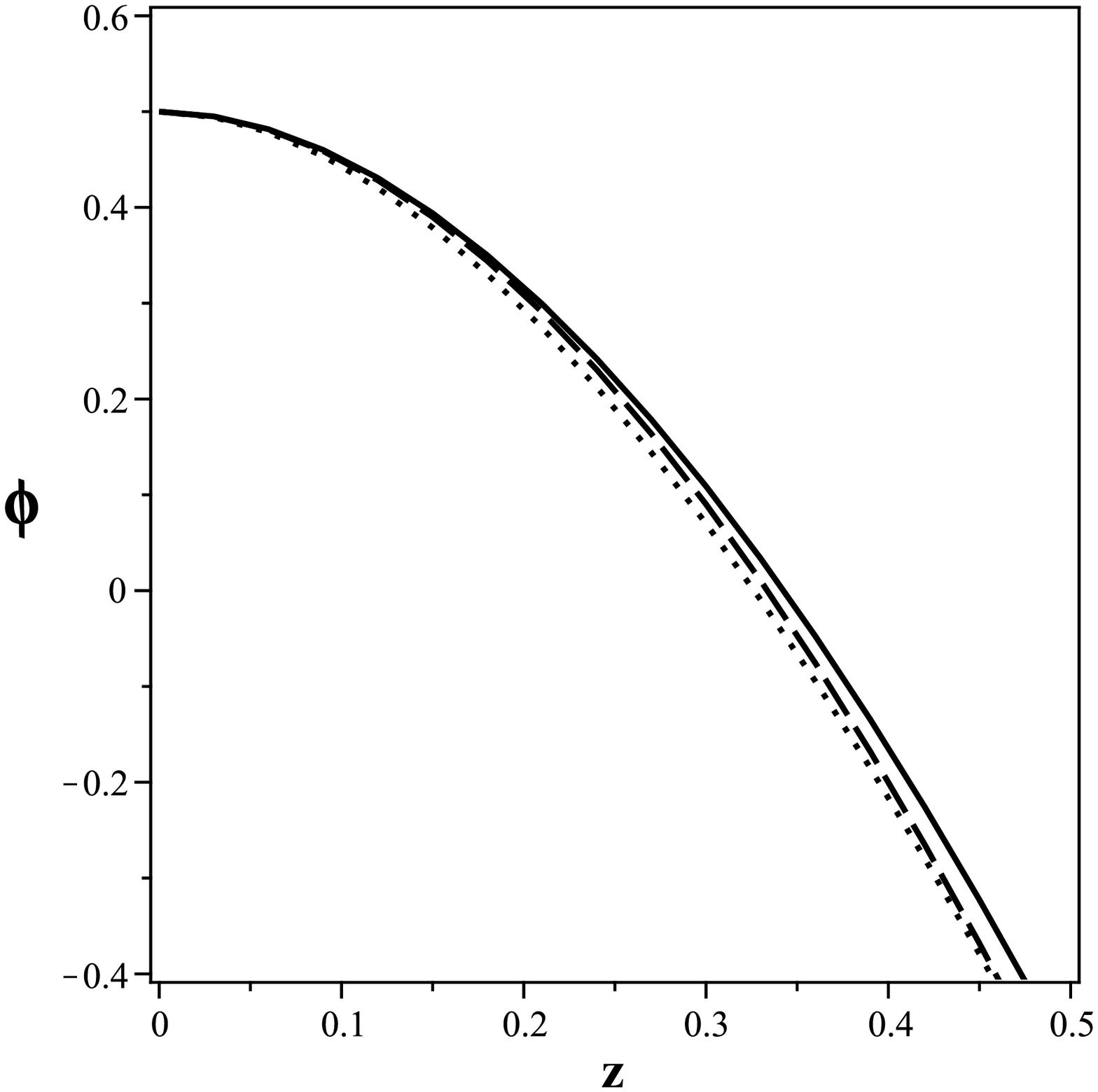}\\
\hspace{1cm}Fig. 5: \, Graphs for the reconstructed $f$ and $\phi$ in respect of $z$. The solid, dot and   \\
\hspace{1cm} dash lines represent parametrization 1, 2 and 3 respectively. Initial values are \\
\hspace{1cm} $\phi(0)=0.5$ , $\dot{\phi}(0)=-0.02$, $\psi(0)=0.5$ , $\dot{\psi}(0)=0.01$ and $f(0)=0.5$.\\
\end{tabular*}\\

One then can reconstruct $f(\phi)$, which is plotted in Fig. 6. The exponential behavior of the reconstructed $f(\phi)$with respected to $\phi$ in particular for the second parametrization is compatible with the one initially assumed in the numerical calculations.\\

\begin{tabular*}{2.5 cm}{cc}
\includegraphics[scale=.3]{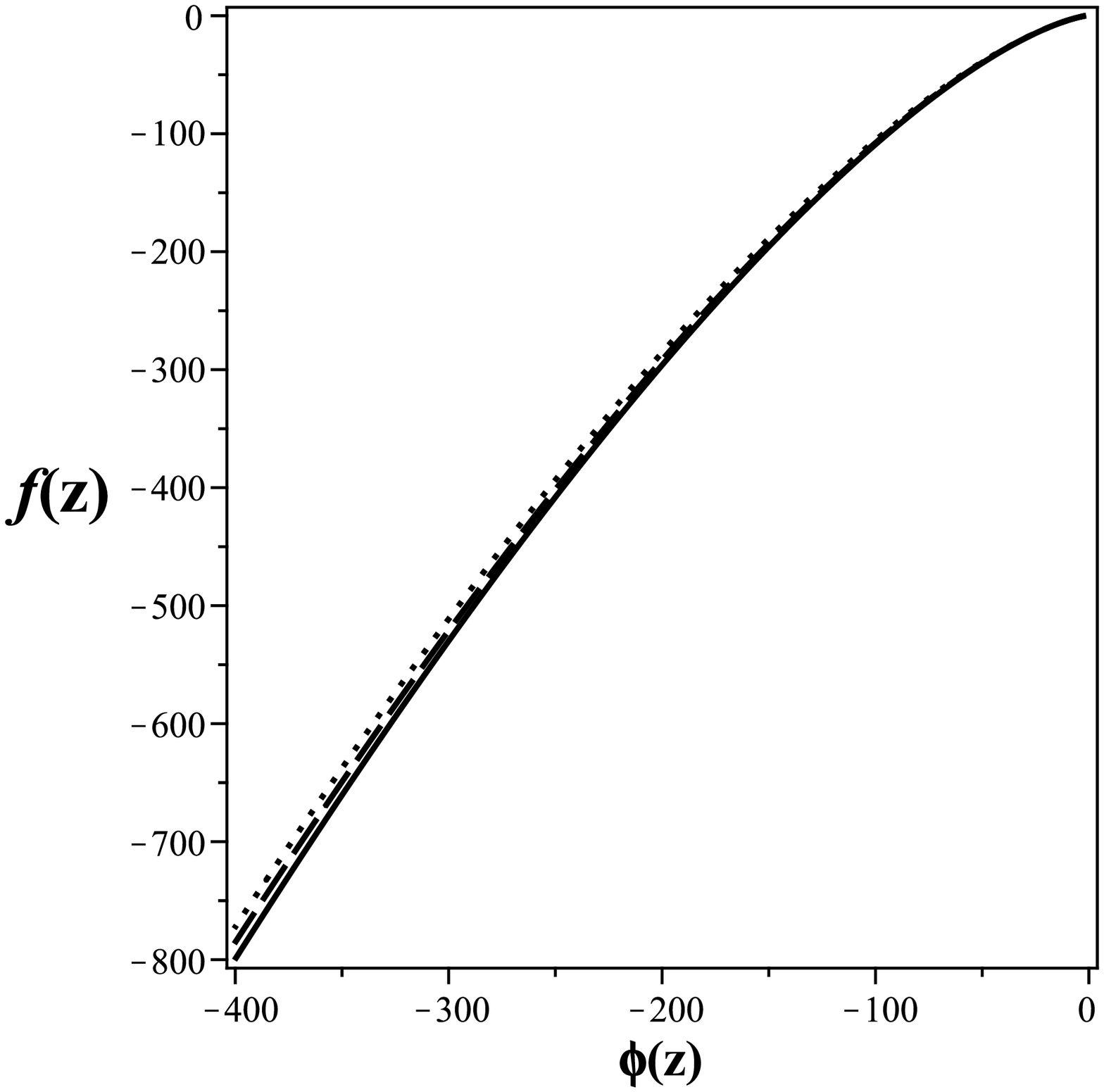} \\
\hspace{1cm}Fig. 6: \, Graphs for the reconstructed $f$ in respect of $\phi$. The solid, dot and   \\
\hspace{1cm} dash lines represent parametrization 1, 2 and 3 respectively. Initial values are \\
\hspace{1cm} $\phi(0)=0.5$ , $\dot{\phi}(0)=-0.02$, $\psi(0)=0.5$ , $\dot{\psi}(0)=0.01$ and $f(0)=0.5$.\\
\end{tabular*}\\

\section{Summary and Conclusion}

In this paper , we consider a local scalar-tensor formulation of non-local gravity as a simple modified model
 characterized by two scalar fields $\phi$ and $\psi$
and function $f(\phi)$ which can be viewed as scalar potential in the model. Analytical study of the solution shows that under special condition,
the universe may go through a transition from quintessence to phantom phase which is also supported by numerical analysis.\\
 In analytic studying of the dynamics of the EoS parameter we achieve the constraints that one has to impose on the scalar fields and their first and second derivatives in order to have phantom crossing. In numerical approach, the EoS parameter crosses $\omega=-1$ for $t>0$. We also find that universe undergoes a bounce, i.e contracts, reaches a minimum radius and then expands.

  Finally,
 we reconstruct our cosmological parameters, such as the EoS parameter, deceleration parameter and potential function $f(\phi)$. In general, different cosmological models can be observationally differentiate in terms of the potential function of the dynamical system. In here, in the construction of the deceleration parameter $q(z)$, it is found that the strongest evidence of acceleration occurs about redshift $z\sim 0.2-0.39$ \cite{c23-02}. Also for the EoS parameter, it is found that $\omega(z)$ crossing to be around redshift $z\sim 0.2$ \cite{c23-02}. This suggest that all three parametrization are suitable for small $z$ and in particular the second one is remarkably close to the observational data. We then reconstructed the potential function of the dynamical system using three forms of parametrization by fitting the model to the observational. The behavior of $f(\phi)$ is similar to the initially assumed exponential form.

\end{document}